\documentclass[12pt]{article}
\pdfoutput=1

\usepackage{amsmath}
\usepackage{amssymb}
\usepackage{epsfig}
\usepackage{epstopdf}
\usepackage{latexsym}
\usepackage{color}
\numberwithin{equation}{section}
\usepackage[
      colorlinks=false,
      linkcolor=darkblue,  
      urlcolor=blue,    
      filecolor=blue,     
      citecolor=red,
linktocpage=true,
      pdfstartview=FitV,
      bookmarksopen=true    
      ]{hyperref}

\begin{document}

\begin{titlepage}

\centerline{\Huge \rm } 
\bigskip
\centerline{\Huge \rm Uplifting supersymmetric $AdS_6$ black holes}
\bigskip
\centerline{\Huge \rm to type II supergravity}
\bigskip
\bigskip
\bigskip
\bigskip
\bigskip
\bigskip
\centerline{\rm Minwoo Suh}
\bigskip
\centerline{\it Department of Physics, Kyungpook National University, Daegu 41566, Korea}
\bigskip
\centerline{\tt minwoosuh1@gmail.com} 
\bigskip
\bigskip
\bigskip
\bigskip
\bigskip
\bigskip
\bigskip
\bigskip
\bigskip
\bigskip
\bigskip
\bigskip

\begin{abstract}
\noindent Employing uplift formulae, we uplift supersymmetric $AdS_6$ black holes from $F(4)$ gauged supergravity to massive type IIA and type IIB supergravity. In massive type IIA supergravity, we obtain supersymmetric $AdS_6$ black holes asymptotic to the Brandhuber-Oz solution. In type IIB supergravity, we obtain supersymmetric $AdS_6$ black holes asymptotic to the non-Abelian T-dual of the Brandhuber-Oz solution. For the uplifted black hole solutions, we calculate the holographic entanglement entropy. In massive type IIA supergravity, it precisely matches the Bekenstein-Hawking entropy of the black hole solutions.
\end{abstract}

\vskip 5cm

\flushleft {August, 2019}

\end{titlepage}

\tableofcontents

\section{Introduction and conclusions}

Recently we are observing progress of the AdS/CFT correspondence, \cite{Maldacena:1997re},  in higher dimensions, $e.g.$, in AdS$_7$/CFT$_6$ and AdS$_6$/CFT$_5$. For supersymmetric $AdS_6$ solutions of string/M-theory, the only known solution was the near horizon limit of the D4-D8 brane system, known as the Brandhuber-Oz solution of massive type IIA supergravity \cite{Brandhuber:1999np} which is dual to 5d superconformal field theories in \cite{Ferrara:1998gv, Seiberg:1996bd, Morrison:1996xf, Intriligator:1997pq}. Later, it was shown to be the unique supersymmetric $AdS_6$ solution of massive type IIA supergravity \cite{Passias:2012vp}. After the discovery of the non-Abelian T-dual of the Brandhuber-Oz solution in type IIB supergravity in \cite{Lozano:2012au}, and derivations of the supersymmetry equations of general $AdS_6$ solutions of type IIB supergravity in \cite{Apruzzi:2014qva, Kim:2015hya, Kim:2016rhs}, an infinite family of $AdS_6$ solutions was discovered in \cite{DHoker:2016ujz, DHoker:2016ysh, DHoker:2017mds, DHoker:2017zwj}. See also \cite{Apruzzi:2018cvq}. Although the dual field theory of the non-Abelian T-dual of the Brandhuber-Oz solutions is still unclear \cite{Lozano:2013oma, Lozano:2018pcp}, the infinite family of $AdS_6$ solutions was shown to be dual of five-brane web theories, $e.g.$, 5d $T_N$ theories, \cite{Gutperle:2017tjo, Bergman:2018hin, Fluder:2018chf, Uhlemann:2019ypp}.

We can also study the AdS$_6$/CFT$_5$ correspondence from six dimensions, $i.e.$, from $F(4)$ gauged supergravity, \cite{Romans:1985tw}. Any solution of $F(4)$ gauged supergravity can be uplifted to a solution of massive type IIA, \cite{Cvetic:1999un}, and type IIB supergravity, \cite{Jeong:2013jfc, Hong:2018amk, Malek:2018zcz}. Moreover, more recently, uplift formula for $F(4)$ gauged supergravity coupled to arbitrary number of vector multiplets to type IIB supergravity has been derived in \cite{Malek:2019ucd}. To the best of our knowledge, it is the first uplift formula for gauged supergravity with $arbitrary$ number of matter multiplets. They also showed that uplifts are only possible for $F(4)$ gauged supergravity coupled $up$ $to$ three vector mutiplets.  Due to the discovery of the infinite family of supersymmetric $AdS_6$ solutions of type IIB supergravity, \cite{DHoker:2016ujz, DHoker:2016ysh, DHoker:2017mds, DHoker:2017zwj}, and the developments in exceptional field theory, \cite{Malek:2017njj, Malek:2018zcz, Malek:2019ucd}, these recent progress in uplift formulae was able.

In this paper, we employ the uplift formulae in \cite{Cvetic:1999un} and \cite{Malek:2019ucd} to some solutions of $F(4)$ gauged supergravity, and obtain solutions of massive type IIA and type IIB supergravity. For this purpose, we choose supersymmetric $AdS_6$ black hole solutions from $F(4)$ gauged supergravity in \cite{Suh:2018tul}. Analogous to the microscopic entropy counting, \cite{Benini:2015noa, Benini:2015eyy}, of supersymmetric $AdS_4$ black holes, \cite{Cacciatori:2009iz, DallAgata:2010ejj, Hristov:2010ri}, the Bekenstein-Hawking entropy of $AdS_6$ black holes, \cite{Suh:2018tul}, was successfully counted by the topologically twisted index, \cite{Hosseini:2018uzp}, of $5d$ $USp(2N)$ gauge theories, \cite{Seiberg:1996bd, Morrison:1996xf, Intriligator:1997pq}. See also \cite{Crichigno:2018adf}, \cite{Suh:2018qyv}, and \cite{Kim:2019fsg}. More general $AdS_6$ black holes in matter coupled $F(4)$ gauged supergravity were found in \cite{Hosseini:2018usu} and further studied in \cite{Suh:2018szn}.

The supersymmetric $AdS_6$ black hole solutions are asymptotic to the supersymmetric $AdS_6$ fixed point and have horizon of $AdS_2\times\Sigma_{\frak{g}_1}\times\Sigma_{\frak{g}_2}$, where $\frak{g}_1>1$ and $\frak{g}_2>1$ are genus of the Riemann surfaces, \cite{Suh:2018tul, Hosseini:2018usu, Suh:2018szn}. The supersymmetric $AdS_6$ fixed point of $F(4)$ gauged supergravity is universal, $i.e.$, it uplifts to i) the Brandhuber-Oz solution of massive type IIA supergravity, \cite{Brandhuber:1999np}, ii) the non-Abelian T-dual of the Brandhuber-Oz solution of type IIB supergravity, \cite{Lozano:2012au}, and iii) the infinite family of $AdS_6$ solutions of type IIB supergravity, \cite{DHoker:2016ujz, DHoker:2016ysh, DHoker:2017mds, DHoker:2017zwj}. For the solutions uplifted to massive type IIA supergravity, as the Brandhuber-Oz solution is the unique supersymmetric $AdS_6$ solution of massive type IIA supergravity, the uplifted solutions are automatically asymptotic to i) the Brandhuber-Oz solution. On the other hand, for the solutions uplifted to type IIB supergravity, solutions can be asymptotic to ii) the non-Abelian T-dual of the Brandhuber-Oz solution or iii) the infinite family of $AdS_6$ solutions. In the uplift formulae in \cite{Malek:2018zcz, Malek:2019ucd}, it is determined by choosing holomorphic functions, $\mathcal{A}_\pm(z)$ where $z$ is a complex coordinate on a Riemann surface, \cite{DHoker:2016ujz, DHoker:2016ysh, DHoker:2017mds, DHoker:2017zwj}. For the solutions asymptotic to the non-Abelian T-dual of the Brandhuber-Oz solutions, the holomorphic functions are given by
\begin{equation}
\mathcal{A}_\pm\,=\,\frac{1}{216}z^3{\mp}\frac{i}{4}z-\frac{i}{108}\,.
\end{equation}
It was first obtained in \cite{Hong:2018amk} and rediscovered in \cite{Lozano:2018pcp}. For the solutions asymptotic to the infinite family of $AdS_6$ solutions, \cite{DHoker:2016ujz, DHoker:2016ysh, DHoker:2017mds, DHoker:2017zwj}, the holomorphic functions are given by
\begin{equation}
\mathcal{A}_\pm\,=\,\mathcal{A}_\pm^0+\sum_{l=1}^LZ_\pm^l\ln(z-r_l)\,,
\end{equation}
where
\begin{equation}
Z_\pm^l\,=\,\sigma\sum_{n=1}^{L-2}(r_l-s_n)\sum_{k\ne{l}}^L\frac{1}{r_l-r_k}\,,
\end{equation}
and $r_l$ and $s_n$ are the positions of the zeros and poles on the Riemann surfaces, respectively, and $\mathcal{A}_\pm^0$ and $\sigma$ are complex constants. However, the infinite family of $AdS_6$ solutions is not able to be evaluated by elementary functions, and either the uplift of the black holes asymptotic to the infinite family of $AdS_6$ solutions. Hence, we restrict ourselves to the uplifts of supersymmetric black holes asymptotic to i) the Brandhuber-Oz solutions of massive type IIA supergravity and ii) the non-Abelian T-dual of the Brandhuber-Oz solutions of type IIB supergravity, and $not$ consider iii) the infinite family of supersymmetric $AdS_6$ solutions of type IIB supergravity.{\footnote{Schematic geometry of black holes asymptotic to the infinite family of supersymmetric $AdS_6$ solutions of type IIB supergravity was considered for entropy counting in \cite{Fluder:2019szh}.}} 

By employing the uplift formula to type IIB supergravity in \cite{Malek:2019ucd}, it will be interesting to uplift the supersymmetric $AdS_6$ black holes from matter coupled $F(4)$ gauged supergravity in \cite{Hosseini:2018usu, Suh:2018szn}. However, there is no consistent truncation to matter coupled $F(4)$ gauged supergravity from solutions asymptotic to the non-Abelian T-dual of the Brandhuber-Oz solution, \cite{Malek:2019ucd}.

For the uplifted black hole solutions, we calculate the holographic entanglement entropy, \cite{Ryu:2006bv, Ryu:2006ef}. For the solutions in massive type IIA supergravity, the holographic entanglement entropy precisely matches the Bekenstein-Hawking entropy of the black hole solutions obtained in \cite{Suh:2018tul, Suh:2018qyv} which is microscopically counted by topologically twisted index of $5d$ $USp(2N)$ gauge theories, \cite{Hosseini:2018uzp}. For the solutions in type IIB supergravity, if the field theory dual of the non-Abelian T-dual of the Brandhuber-Oz solution could be identified, it would be interesting to compare the holographic entanglement entropy we obtain here with the microscopic results.

In section 2, we review the uplift formula for $F(4)$ gauged supergrvity to massive type IIA supergravity, and employ the uplift formula to obtain supersymmetric $AdS_6$ black holes asymptotic to the Brandhuber-Oz solutions of massive type IIA supergravity. In section 3, we review the uplift formula for $F(4)$ gauged supergravity to type IIB supergravity, and employ the uplift formula to obtain supersymmetric $AdS_6$ black holes asymptotic to the non-Abelian T-dual of the Brandhuber-Oz solutions of type IIB supergravity. Some technical details are relegated in appendices.

\section{Supersymmetric $AdS_6$ black holes of massive type IIA supergravity}

\subsection{Massive type IIA from $F(4)$ gauged supergravity}

We review the uplift formula for $F(4)$ gauged supergravity to massive type IIA supergravity in  \cite{Cvetic:1999un}.

We introduce the bosonic field content of $F(4)$ gauged supergravity, \cite{Romans:1985tw}, in the conventions of \cite{Cvetic:1999un}.{\footnote{The couplings and fields in the uplift formula for pure $F(4)$ gauged supergravity to massive type IIA supergravity in \cite{Cvetic:1999un} are related to the ones of \cite{Romans:1985tw} by
\begin{align}
\tilde{g}\,=&\,2g\,, \qquad X\,=\,e^{-\frac{\phi}{2\sqrt{2}}}\,=\,e^{\frac{\tilde{\phi}}{\sqrt{2}}}\,, \notag \\
\tilde{A}^I_\mu\,=&\,\frac{1}{2}A^I_\mu\,, \qquad \tilde{A}_\mu\,=\,\frac{1}{2}A_\mu\,, \qquad \tilde{B}_{\mu\nu}\,=\,\frac{1}{2}B_{\mu\nu}\,,
\end{align}
where the tilded ones are of \cite{Romans:1985tw}.}} There are the metric, the real scalar field, $\phi$, an $SU(2)$ gauge field, $A^I$, $I\,=\,1,\,2,\,3$, a $U(1)$ gauge field, $A$, and a two-form gauge potential, $B$. Their field strengths are respectively defined by
\begin{align}
\tilde{F}_{(2)}^I\,=&\,dA^I+\frac{1}{2}g\epsilon^{IJK}A_J\wedge{A}_K\,, \notag \\
\tilde{F}_{(2)}\,=&\,dA+\frac{2}{3}gB\,, \notag \\
\tilde{F}_{(3)}\,=&\,dB\,.
\end{align}
There are two parameters, the $SU(2)$ gauge coupling, $g$, and the mass parameter of the two-form field, $m$. When $g>0$, $m>0$ and $g\,=\,3m/2$, the theory admits a unique supersymmetric $AdS_6$ fixed point. At the fixed point, all the fields are vanishing except the $AdS_6$ metric.

The uplift formula for $F(4)$ gauged supergravity to massive type IIA supergravity was obtained in \cite{Cvetic:1999un}. The uplifted metric and the dilaton field are given by, respectively,
\begin{align} \label{mIIAmet}
ds^2\,=&\,X^{1/8}\sin^{1/12}\xi\left(\Delta^{3/8}ds_6^2+\frac{2}{g^2}\Delta^{3/8}X^2d\xi^2+\frac{1}{2g^2}\frac{\cos^2\xi}{\Delta^{5/8}X}ds_{\tilde{S}^3}^2\right)\,, \\ \label{mIIAdil}
e^{\Phi}\,=&\,\frac{\Delta^{1/4}}{X^{5/4}\sin^{5/6}\xi}\,,
\end{align}
and the two-, three-, and four-form fluxes are given by, respectively,{\footnote{We suspect a typographical error in \cite{Cvetic:1999un}. We changed the sign of $*_6\tilde{F}_{(2)}$ term in $F_{(4)}$ from \cite{Cvetic:1999un}.}
\begin{align} \label{mIIAtwo}
F_{(2)}\,=\,&\frac{1}{\sqrt{2}}\sin^{2/3}\xi\tilde{F}_{(2)}\,, \\ \label{mIIAthree}
H_{(3)}\,=\,&\sin^{2/3}\xi\tilde{F}_{(3)}+\frac{\cos\xi}{g\sin^{1/3}\xi}\tilde{F}_{(2)}\wedge{d}\xi\,, \\ \label{mIIAfour}
F_{(4)}\,=\,&-\frac{\sqrt{2}}{6}\frac{U\sin^{1/3}\xi\cos^3\xi}{g^3\Delta^2}d\xi\wedge{vol}_{\tilde{S}^3}-\sqrt{2}\frac{\sin^{4/3}\xi\cos^4\xi}{g^3\Delta^2X^3}dX\wedge{vol}_{\tilde{S}^3} \notag \\
&-\sqrt{2}\frac{X^4\sin^{1/3}\xi\cos\xi}{g}*_6\tilde{F}_{(3)}\wedge{d}\xi+\frac{1}{\sqrt{2}}\frac{\sin^{4/3}\xi}{X^2}*_6\tilde{F}_{(2)} \notag \\
&+\frac{1}{\sqrt{2}}\frac{\sin^{1/3}\xi\cos\xi}{g^2}\tilde{F}_{(2)}^I\wedge{h}^I\wedge{d}\xi-\frac{1}{4\sqrt{2}}\frac{\sin^{4/3}\xi\cos^2\xi}{g^2\Delta{X}^3}\tilde{F}_{(2)}^I\wedge{h}^J\wedge{h}^K\epsilon_{IJK}\,.
\end{align}
We employ the metric and the volume form on the gauged three-sphere by
\begin{align}
ds_{\tilde{S}^3}^2\,=&\,\sum^3_{I=1}\left(\sigma^I-gA^I\right)^2\,, \notag \\
vol_{\tilde{S}^3}\,=&\,h_1\wedge{h_2}\wedge{h_3}\,,
\end{align}
where
\begin{align}
h^I\,=\,\sigma^I-gA^I\,,
\end{align}
and $\sigma^I$, $I\,=\,1,\,2,\,3$, are the $SU(2)$ left-invariant one-forms which satisfy
\begin{equation}
d\sigma^I\,=\,-\frac{1}{2}\epsilon_{IJK}\sigma^J\wedge\sigma^K\,.
\end{equation}
A choice of the left-invariant one-forms is
\begin{align}
\sigma^1\,=&\,-\sin\alpha_2\cos\alpha_3d\alpha_1+\sin\alpha_3d\alpha_2\,, \notag \\
 \sigma^2\,=&\,\sin\alpha_2\sin\alpha_3d\alpha_1+\cos\alpha_3d\alpha_2\,,  \notag \\
 \sigma^3\,=&\,\cos\alpha_2d\alpha_1+d\alpha_3\,.
\end{align}
We also defined quantities,{\footnote{For the scalar field, $\phi$, we follow the normalization of  \cite{Romans:1985tw} than \cite{Cvetic:1999un}.}}
\begin{align}
X\,=&\,e^{\frac{\phi}{\sqrt{2}}}\,, \notag \\
\Delta\,=&\,X\cos^2\xi+X^{-3}\sin^2\xi\,, \notag \\
U\,=&\,X^{-6}\sin^2\xi-3X^2\cos^2\xi+4X^{-2}\cos^2\xi-6X^{-2}\,.
\end{align}

\subsection{Supersymmetric $AdS_6$ black holes}

We uplift the recently obtained supersymmetric $AdS_6$ black holes of $F(4)$ gauged supergravity in \cite{Suh:2018tul} to massive type IIA supergravity.{\footnote {See \cite{Dibitetto:2018gtk} for uplift of another $AdS_2$ solution in $F(4)$ gauged supergravity to massive type IIA supergravity.}} 

We first present the six-dimensional solutions of \cite{Suh:2018tul} in the conventions of \cite{Cvetic:1999un}. The solutions are given by the metric,
\begin{equation} \label{mabhmet}
ds_6^2\,=\,e^{2f(r)}\left(-dt^2+dr^2\right)+e^{2g_1(r)}\left(d\theta_1^2+\sinh^2\theta_1d\phi_1^2\right)+e^{2g_2(r)}\left(d\theta_2^2+\sinh^2\theta_1d\phi_2^2\right)\,,
\end{equation}
which is asymptotically $AdS_6$ and has a horizon of $AdS_2\,\times\,\Sigma_{\frak{g}_1}\,\times\,\Sigma_{\frak{g}_2}$ with $\frak{g}_1>1$ and $\frak{g}_2>1$. Only a component of the $SU(2)$ gauge field is non-trivial,
\begin{equation} \label{mabhgauge}
A^3\,=\,2\Big[a_1\cosh\theta_1{d}\phi_1+a_2\cosh\theta_2{d}\phi_2\Big]\,,
\end{equation}
where the magnetic charges, $a_1$ and $a_2$, are constant. The twist condition on the magnetic charges is
\begin{equation} \label{matwist}
a_1\,=\,\frac{1}{2}\left[-\frac{k}{\lambda{g}}\right]\,, \qquad a_2\,=\,\frac{1}{2}\left[-\frac{k}{\lambda{g}}\right]\,,
\end{equation}
where $k\,=\,-1$ for $\frak{g}_1>1$ and $\frak{g}_2>1$ and $\lambda\,=\,\pm1$. There is also a two-form field,
\begin{equation} \label{mabhtwo}
B_{tr}\,=\,2\left[-\frac{2}{m^2}a_1a_2e^{\sqrt{2}\phi+2f-2g_1-2g_2}\right]\,,
\end{equation}
and the three-form field strength of the two-form field vanishes identically. There is also a non-trivial real scalar field, $X(r)\,=\,e^{\frac{\phi}{\sqrt{2}}}$, but the $U(1)$ gauge field is vanishing. To match the conventions of \cite{Cvetic:1999un}, there are additional overall factors of 2 to the solutions of \cite{Suh:2018tul} in \eqref{mabhgauge}, \eqref{mabhtwo} and 1/2 in \eqref{matwist}. Also we set $m\,=\,2g/3$. We find the $AdS_2$ horizon for the $\Sigma_{\frak{g}_1}\times\Sigma_{\frak{g}_2}$ background with $\frak{g}_1>1$ and $\frak{g}_2>1$,
\begin{equation} \label{ads2sol}
e^f\,=\,\frac{2^{1/4}}{(2g)^{3/4}m^{1/4}}\frac{1}{r}\,, \qquad e^{g_1}\,=\,e^{g_2}\,=\,\frac{2^{3/4}}{(2g)^{3/4}m^{1/4}}\,, \qquad e^{\frac{\phi}{\sqrt{2}}}\,=\,\frac{2^{1/4}m^{1/4}}{(2g)^{1/4}}\,.
\end{equation}
In order to have unit radius for $AdS_6$, $L\,=\,1$, we set $m\,=\,\sqrt{2}$.

As the uplifted solution is simply the uplift formula in \eqref{mIIAmet}, \eqref{mIIAdil}, \eqref{mIIAtwo}, \eqref{mIIAthree}, and \eqref{mIIAfour} with the six-dimensional fields in \eqref{mabhmet}, \eqref{mabhgauge}, and \eqref{mabhtwo}, it is unnecessary to present them in detail here. The full geometry is interpolating between the asymptotic $AdS_6$ fixed point, which is the Brandhuber-Oz solution, and the near horizon geometry, $AdS_2\times\Sigma_{\frak{g}_1}\times\Sigma_{\frak{g}_2}\times\tilde{S}^3\times{S}^1$, where $S^1$ is parametrized by the coordinate, $0\le\xi\le\pi/2$.

By employing the supersymmetry equations of $F(4)$ gauged supergravity, we explicitly checked that the uplifted solution solves the equations of motion of massive type IIA supergravity. We present the supersymmetry equations of $F(4)$ gauged supergravity in appendix A and the equations of motion of massive type IIA supergravity in appendix B.1.

\subsection{Holographic entanglement entropy}

The black hole solution interpolates between the boundary, $AdS_6\times\tilde{S}^3\times{S}^1$, and the horizon, $AdS_2\times\Sigma_{\frak{g}_1}\times\Sigma_{\frak{g}_2}\times\tilde{S}^3\times{S}^1$. The holographic entanglement entropy, \cite{Ryu:2006bv, Ryu:2006ef}, on the boundary was calculated and matched to the five-sphere free energy of the dual field theory, which is $5d$ $USp(2N)$ gauge theory, \cite{Jafferis:2012iv}. In this section, we calculate the holographic entanglement entropy on the horizon and find that it is precisely the Bekenstein-Hawking entropy of the black hole solutions.

We begin by switching the solution to be in string frame, 
\begin{equation}
g_{MN}^{\text{string}}\,=\,e^{\Phi/2}g_{MN}^{\text{Einstein}}\,,
\end{equation}
and rescale by the conventions of \cite{Brandhuber:1999np, Bergman:2012kr, Bergman:2012qh}, in particular, by (2), (3), and (4) of \cite{Bergman:2012qh}. The metric and the dilaton are given by{\footnote{We suspect that the metric in (4.1) of \cite{Jafferis:2012iv} is missing the factor of $\Omega^2$ and the expression of $L^4$ in (4.3) of \cite{Jafferis:2012iv} is not correct. However, they conspire to cancel each other and the final answer of entanglement entropy is correct.}}
\begin{equation} \label{rscmet}
ds^2\,=\,\Omega^2\frac{\Delta^{1/8}}{X^{1/8}\sin^{1/3}\xi}\left(L^2\Delta^{3/8}ds_6^2+\frac{4L^2}{9}\left(\Delta^{3/8}X^2d\xi^2+\frac{1}{4}\frac{\cos^2\xi}{\Delta^{5/8}X}ds_{\tilde{S}^3}^2\right)\right)\,,
\end{equation}
\begin{equation}
e^{-2\Phi}\,=\,\frac{4L^2}{9}\Omega^{-10}\sin^{5/3}\xi\,,
\end{equation}
where we introduce
\begin{equation}
\Omega\,=\,\left(\frac{3}{2}\mathfrak{m}\right)^{-1/6}\,, \qquad
L^4\,=\,\frac{3^{8/3}\pi{N}}{2^{2/3}\mathfrak{m}^{1/3}}\,, \qquad
\mathfrak{m}\,=\,\frac{8-N_f}{2\pi}\,,
\end{equation}
and $\mathfrak{m}$ is the Romans mass. The coupling constant, $g$, is related to the $AdS_6$ radius, $L$, by
\begin{equation}
\frac{2}{g^2}\,=\,\frac{4L^2}{9}\,.
\end{equation}
We also collect some relations of the Newton's gravitational constant, $G_N^{(10)}$, and the string length, $l_s$, which we take to be $l_s\,=\,1$,{\footnote{See, for example, footnote 3 in \cite{Brandhuber:1999np}.}}
\begin{equation}
16\pi{G}_N^{(10)}\,=\,2\kappa_{10}^2\,, \qquad \kappa_{10}\,=\,8\pi^{7/2}l_s^4\,,
\end{equation}
and, therefore, we obtain
\begin{equation}
\frac{1}{4G_N^{(10)}}\,=\,\frac{2}{(2\pi)^6l_s^8}\,.
\end{equation}

Now we readily calculate the holographic entanglement entropy on the horizon of the black holes. For the gravitational theory with a dilaton field, the holographic entanglement entropy is given by, \cite{Klebanov:2007ws}, 
\begin{equation}
S_{\text{EE}}\,=\,\frac{1}{4G_N^{(10)}}\int{d^8x}e^{-2\Phi}\sqrt{\gamma}\,,
\end{equation}
where $\gamma$ is determinant of the induced metric of co-dimension two entangling surface in string frame. For the metric in \eqref{rscmet}, the determinant of the induced metric is
\begin{equation}
\sqrt{\gamma}\,=\,\left(\frac{4\pi}{3(8-N_f)}\right)^{4/3}\frac{16L^4}{81}e^{2g_1+2g_2}X^{-5/2}\Delta^{1/2}\cos^3\xi\sin^{-4/3}\xi\sinh\theta_1\sinh\theta_2\sin\alpha_2\,.
\end{equation}
We compute the holographic entanglement entropy on the horizon of $AdS_2\times\Sigma_{\frak{g}_1}\times\Sigma_{\frak{g}_2}\times\tilde{S}^3\times{S}^1$. Unlike computing entanglement entropy on the asymptotically $AdS_6$ boundary in \cite{Jafferis:2012iv}, the minimal surface is just a point for $AdS_2$, \cite{Azeyanagi:2007bj}. In this case, the holographic entanglement entropy computes the entanglement between CFTs on two boundaries of $AdS_2$ space. Then, the holographic entanglement entropy is obtained to be
\begin{align} \label{eentropy}
S_{\text{EE}}\,=&\,\frac{3N^{5/2}}{\pi^3\sqrt{2(8-N_f)}}e^{2g_1}vol_{\Sigma_{\mathfrak{g}_1}}e^{2g_2}vol_{\Sigma_{\mathfrak{g}_2}}vol_{S^3}\int_0^{\pi/2}{d}\xi\cos^3\xi\sin^{1/3}\xi\,, \notag \\
=&\,\frac{8\sqrt{2}\pi(\mathfrak{g}_1-1)(\mathfrak{g}_2-1)N^{5/2}}{5\sqrt{8-N_f}}\,,
\end{align}
where we use $e^{2g_1+2g_2}\,=\,2/3^3$ from \eqref{ads2sol} and that the areas are given by
\begin{equation}
vol_{\Sigma_{\mathfrak{g}\ne1}}\,=\,4\pi(\mathfrak{g}-1)\,, \qquad vol_{S^3}\,=\,2\pi^2\,.
\end{equation}
The holographic entanglement entropy in \eqref{eentropy} precisely matches the Bekenstein-Hawking entropy of the black hole calculated in \cite{Suh:2018tul, Suh:2018qyv}. It is microscopically counted by topologically twisted index of $5d$ $USp(2N)$ gauge theories, \cite{Hosseini:2018uzp}. See also \cite{Crichigno:2018adf}.

\vspace{3.4cm}

\section{Supersymmetric $AdS_6$ black holes of type IIB supergravity}

\subsection{Type IIB from $F(4)$ gauged supergravity}

We review the uplift formula for $F(4)$ gauged supergravity to type IIB supergravity in \cite{Malek:2019ucd}.

The general supersymmetric $AdS_6$ solutions of \cite{DHoker:2016ujz, DHoker:2016ysh, DHoker:2017mds, DHoker:2017zwj} are specified by two holomorphic functions, $\mathcal{A}_\pm$. For the uplift formula in \cite{Malek:2018zcz, Malek:2019ucd}, the functions, $\mathcal{A}_\pm$, were redefined by two holomorphic functions,
\begin{equation}
\mathcal{A}_\pm\,=\,if^1{\pm}f^2\,,
\end{equation}
where
\begin{equation}
f^\alpha\,=\,-p^\alpha+ik^\alpha\,,
\end{equation}
and $p^\alpha$ and $k^\alpha$ are real functions. The $SL(2,\mathbb{R})$ indices, $\alpha,\,\beta\,=\,1,\,2$, are raised and lowered by $\epsilon_{12}\,=\,\epsilon^{12}\,=\,1$,
\begin{equation}
f^\alpha\,=\epsilon^{\alpha\beta}f_\beta\,, \qquad f_\alpha\,=\,f^\beta\epsilon_{\beta\alpha}\,.
\end{equation}
We also define quantities,{\footnote {The function, $\gamma$, was denoted by $r$ in \cite{Malek:2018zcz, Malek:2019ucd}.}
\begin{align}
d\gamma\,=&\,-p_{\alpha}dk^{\alpha}\,, \\
|dk|\,=&\,\frac{1}{2}\partial_\alpha{k}_\beta\partial^\alpha{k}^\beta\,, \\
\overline{\Delta}\,=&\,3\gamma|dk|^2+X^4|dk|p_{\gamma}p_{\delta}\partial_{\sigma}k^\gamma\partial^{\sigma}p^\delta\,,
\end{align}
where $X$ is the real scalar field of $F(4)$ gauged supergravity,
\begin{equation}
X\,=\,e^{\frac{\phi}{\sqrt{2}}}\,.
\end{equation}

We introduce the bosonic field content of $F(4)$ gauged supergravity, \cite{Romans:1985tw}, reparametrized to match conventions of the uplift formula in \cite{Malek:2018zcz, Malek:2019ucd}. There are the metric, the real scalar field, $\phi$, an $SU(2)$ gauge field, $A^A$, $A\,=\,1,\,2,\,3$, a $U(1)$ gauge field, $A^4$, and a two-form gauge potential, $B$. Their field strengths are respectively defined by
\begin{align}
\tilde{F}_{(2)}^A\,=&\,dA^A+\frac{3}{2\sqrt{2}R}\epsilon^{ABC}A_B\wedge{A}_C\,, \notag \\
\tilde{F}_{(2)}^4\,=&\,dA^4+\frac{\sqrt{2}}{R}B\,, \notag \\
\tilde{F}_{(3)}\,=&\,dB\,.
\end{align}
There are two parameters, the $SU(2)$ gauge coupling, $g$, and the mass parameter of the two-form field, $m$. When $g>0$, $m>0$ and $g\,=\,3m$, the theory admits a unique supersymmetric $AdS_6$ fixed point. At the fixed point, all the fields are vanishing except the $AdS_6$ metric.

The uplifted geometry is going to be of the form,
\begin{equation}
M_{1,5}\,\times\,\tilde{S}^2\,\times\,\Sigma\,,
\end{equation}
where $M_{1,5}$ is given by the metric of $F(4)$ gauged supergravity, and $\Sigma$ is the Riemann surface parametrized by a complex coordinate, $z$. For the gauged two-sphere, $\tilde{S}^2$, we introduce real coordinates, $y^A$, $A\,=\,1,\,2,\,3$, their derivatives,
\begin{equation}
Dy^A\,=\,dy^A+\frac{3}{\sqrt{2}R}\epsilon^{ABC}A_By_C\,,
\end{equation}
and their Hodge dual coordinates,
\begin{equation}
\tilde{\theta}_A\,=\,*_{\tilde{S}^2}Dy_A\,=\,\epsilon_{ABC}y^BDy^C\,.
\end{equation}
The metric and the volume form of the gauged two-sphere are, respectively, given by
\begin{align}
ds_{\tilde{S}^2}\,=&\,\delta_{AB}Dy^A\otimes{Dy}^B\,, \notag \\
vol_{\tilde{S}^2}\,=&\,\frac{1}{2}\epsilon_{ABC}y^ADy^B\wedge{Dy}^C\,.
\end{align}

Now we readily present the uplift formula for $F(4)$ gauged supergravity to type IIB supergravity in \cite{Malek:2019ucd}. The metric, the dilaton-axion fields, and the NSNS and RR two-form gauge potentials are respectively obtained from
\begin{align}
ds^2\,=&\,\frac{4\,3^{1/4}c_6\gamma^{1/4}\overline{\Delta}^{1/4}}{|dk|^{1/2}}ds_6^2+\frac{4c_6R^2\gamma^{5/4}|dk|^{3/2}X^2}{3^{3/4}\overline{\Delta}^{3/4}}ds_{\tilde{S}^2}^2+\frac{4c_6R^2\overline{\Delta}^{1/4}}{3^{3/4}\gamma^{3/4}|dk|^{1/2}X^2}dk^\alpha{\otimes}dp_\alpha\,, \notag \\
H_{\alpha\beta}\,=&\,\frac{1}{\sqrt{3\overline{\Delta}}}\left(\frac{X^4|dk|}{\sqrt{\gamma}}p_{\alpha}p_\beta+3\sqrt{\gamma}\partial_\gamma{k}_\alpha\partial^\gamma{p}_\beta\right)\,, \notag \\
C_{(2)}^\alpha\,=&\,-\frac{4c_6R^2}{3}\left(k^\alpha+\frac{X^4\gamma|dk|}{\overline{\Delta}}p_\gamma\partial_\beta{k}^\gamma\partial^\beta{p}^\alpha\right)vol_{\tilde{S}^2} \notag \\ &+2\sqrt{2}c_6RA^A\wedge\left(y_Adk^\alpha+k^\alpha{D}y_A\right)+2\sqrt{2}c_6RA^4\wedge{d}p^\alpha+4c_6Bp^\alpha-3c_6k^\alpha\epsilon_{ABC}y^AA^B\wedge{A}^C\,,
\end{align}
where $R$ and $c_6$ are constants. The five-form flux is obtained from
\begin{equation}
F_{(5)}\,=\,F_{(2,3)}+F_{(3,2)}+F_{(4,1)}\,,
\end{equation}
where
\begin{align}
F_{(2,3)}\,=&\,\frac{8\sqrt{2}c_6^2R^3|dk|}{3}\left(\tilde{F}_{(2)}^A\wedge\tilde{\theta}_A\wedge{vol}_\Sigma+\frac{X^4\gamma|dk|}{\overline{\Delta}}\left({y}_A\tilde{F}_{(2)}^A\wedge{p}_{\alpha}dp^\alpha-\tilde{F}_{(2)}^4\wedge{p}_{\alpha}dk^\alpha\right)\wedge{vol}_{\tilde{S}^2}\right)\,, \notag \\
F_{(3,2)}\,=&\,16c_6^2R^2\left(\frac{\gamma^2|dk|^2}{\overline{\Delta}}\tilde{F}_{(3)}\wedge{vol}_{\tilde{S}^2}+\frac{|dk|}{X^4}*_6\tilde{F}_{(3)}\wedge{vol}_\Sigma\right)\,, \notag \\
F_{(4,1)}\,=&\,8\sqrt{2}c_6^2RX^2\left(-\gamma*_6\tilde{F}_{(2)}^A\wedge{D}y_A+*_6\tilde{F}_{(2)}^4\wedge{p}_{\alpha}dp^\alpha+y_A*_6\tilde{F}_{(2)}^A\wedge{p}_{\alpha}dk^\alpha\right)\,,
\end{align}
and $F_{(4,1)}\,=\,*F_{(2,3)}$.

\subsection{Uplifted solutions}

For the solutions asymptotic to the non-Abelian T-dual of the Brandhuber-Oz solution, the form of the functions, $\mathcal{A}_\pm$, was first obtained in \cite{Hong:2018amk} and rediscovered in \cite{Lozano:2018pcp},
\begin{equation} \label{Afunction}
\mathcal{A}_\pm\,=\,\frac{1}{c_6}\left(\frac{1}{216}z^3{\mp}\frac{i}{4}z-\frac{i}{108}\right)\,,
\end{equation}
where $z$ is a complex coordinate on the Riemann surface. We can obtain explicit expressions for the quantities defined in the previous subsection, $e.g.$,
\begin{equation}
f^1\,=\,-\frac{1}{216ic_6}\left(z^3-2i\right)\,, \qquad f^2\,=\,-\frac{1}{4ic_6}z\,,
\end{equation}
\begin{equation}
p^1\,=\,-\frac{1}{432ic_6}\left(z^3-\bar{z}^3-4i\right)\,, \qquad p^2\,=\,-\frac{1}{8ic_6}\left(z-\bar{z}\right)\,,
\end{equation}
\begin{equation}
k^1\,=\,-\frac{1}{432c_6}\left(z^3+\bar{z}^3\right)\,, \qquad k^2\,=\,-\frac{1}{8c_6}\left(z+\bar{z}\right)\,,
\end{equation}
and
\begin{equation}
\gamma\,=\,\frac{1}{864c_6^2}\left(\frac{1}{8i}\left(z^4-\bar{z}^4\right)+\frac{1}{4i}\left(z\bar{z}^3-\bar{z}z^3\right)+z+\bar{z}\right)\,,
\end{equation}
\begin{equation}
|dk|\,=\,\frac{1}{576ic_6^2}\left(z^2-\bar{z}^2\right)\,.
\end{equation}

For the gauged two-sphere, we introduce angular coordinates,
\begin{equation}
y^1\,=\,\sin\theta\sin\phi\,, \qquad y^2\,=\,\sin\theta\cos\phi\,, \qquad y^3\,=\,-\cos\theta\,.
\end{equation}
In terms of the angular coordinates, a Hodge dual coordinate and the volume form are
\begin{align}
\tilde{\theta}_3\,=&\,-\sin^2\theta{D}\phi\,, \\
vol_{\tilde{S}^2}\,=&\,\sin\theta{d}\theta{D}\phi\,.
\end{align}

For the Riemann surface, we also introduce angular coordinates, $e.g.$, (4.3) of \cite{Hong:2018amk},
\begin{equation} \label{zx1x2}
z\,=\,x^1+ix^2\,,
\end{equation}
where
\begin{equation}
x^1\,=\,\frac{2\tilde{g}^2}{3}\rho\,, \qquad x^2\,=\,\sin^{2/3}\xi\,.
\end{equation}
The metric and the volume form are given, respectively, by
\begin{align}
ds_\Sigma^2\,=&\,dzd\bar{z}\,=\,\frac{4\tilde{g}^4}{9}\left(d\rho^2+\frac{\cos^2\xi}{\tilde{g}^4\sin^{2/3}\xi}d\xi^2\right)\,, \\
vol_{\Sigma}\,=&\,\frac{1}{2}\epsilon_{\alpha\beta}dx^\alpha\wedge{dx}^\beta\,=\,\frac{4\tilde{g}^2}{9}\frac{\cos\xi}{\sin^{1/3}\xi}d\rho\wedge{d}\xi\,.
\end{align}

For the later use we also give explicit expressions of $p^\alpha$ and $k^\alpha$ in terms of the angular coordinates on the Riemann surface,
\begin{align}
k^\alpha\,=&\,\left\{-\frac{1}{2916}\left(4\tilde{g}^6\rho^3-27\tilde{g}^2\rho\sin^{4/3}\xi\right),-\frac{1}{6}\tilde{g}^2\rho\right\}\,, \notag \\
dk^\alpha\,=&\,\left\{-\frac{1}{972}\left(4\tilde{g}^6\rho^2-9\tilde{g}^2\sin^{4/3}\xi\right)d\rho+\frac{1}{81}\tilde{g}^2\rho\cos\xi\sin^{1/3}\xi{d}\xi,-\frac{1}{6}\tilde{g}^2d\rho\right\}\,, \notag \\
p^\alpha\,=&\,\left\{\frac{1}{648}\left(6+3\sin^2\xi-4\tilde{g}^4\rho^2\sin^{2/3}\xi\right),-\frac{1}{4}\sin^{2/3}\xi\right\}\,, \notag \\
dp^\alpha\,=&\,\left\{-\frac{1}{81}\tilde{g}^4\rho\sin^{2/3}\xi{d\rho}+\frac{1}{972}\cos\xi\left(9\sin\xi-\frac{4\tilde{g}^4\rho^2}{\sin^{1/3}\xi}\right)d\xi,-\frac{1}{6}\frac{\cos\xi}{\sin^{1/3}\xi}d\xi\right\}\,.
\end{align}

In addition to the $SU(2)$ gauge coupling, $g$, and the mass parameter of the two-form field, $m$, of $F(4)$ gauged supergravity, we introduced a parameter, $\tilde{g}$, $e.g.$, \cite{Jeong:2013jfc}, by
\begin{equation} \label{gtilde}
\tilde{g}\,=\,\frac{(3g^3m)^{1/4}}{2}\,.
\end{equation}
For the solutions we consider, we will choose the parameters to be
\begin{equation}
g\,=\,3m\,=\,2\sqrt{2}.
\end{equation}
Therefore, we have
\begin{equation}
\tilde{g}\,=\,\sqrt{2}\,.
\end{equation}
Accordingly, we also specify the constants,
\begin{equation}
c_6\,=\,1\,, \qquad R\,=\,\frac{3}{2}\,.
\end{equation}

\subsubsection{Supersymmetric $AdS_6$ fixed point}

As the simplest exercise, we uplift the supersymmetric fixed point of $F(4)$ gauged supergravity to the non-Abelian T-dual of the Brandhuber-Oz solution in type IIB supergravity, \cite{Lozano:2012au}. This uplift was previously done in \cite{Jeong:2013jfc, Hong:2018amk}.

Employing the uplift formula with $\mathcal{A}_\pm$ in \eqref{Afunction}, we obtain the metric,
\begin{equation}
ds^2=\frac{1}{3\sqrt{3}}\left[\frac{F^{1/4}\cos^{3/2}\xi}{\sin^{1/6}\xi}ds_{AdS_6}^2+R^2\left(\frac{4\tilde{g}^4}{9}\frac{X^2\rho^2\tan^{1/2}\xi}{F^{3/4}}ds_{S^2}^2+\frac{F^{1/4}\tan^{1/2}\xi}{X^2}ds_\Sigma^2\right)\right],
\end{equation}
where $F$ is given by
\begin{equation}\label{Ffunction}
F\,=\,4\tilde{g}^4\rho^2\sec^2\xi\sin^{2/3}\xi+X^4\left(1+4\tilde{g}^4\rho^2\sec^4\xi\sin^{8/3}\xi\right)\,.
\end{equation}
Employing an $SL(2,\mathbb{R})$ rotation,
\begin{equation} \label{uuuu}
U\,=\,\left(
\begin{array}{ll}
 \,\,\,\,0 & 1 \\
 -1 & 0
\end{array}
\right)\,,
\end{equation}
to the dilaton-axion field matrix, $H_{\alpha\beta}$, we obtain
\begin{equation} \label{uuhhuu}
U^{-1}HU\,=\,e^\Phi\left(
\begin{array}{ll}
 \, |\tau|^2 & C_{(0)} \\
 C_{(0)} & \,\,\, 1
\end{array}
\right)\,,
\end{equation}
where
\begin{equation}
\tau\,=\,C_{(0)}+ie^{-\Phi}\,.
\end{equation}
The dilaton and axion fields are obtained to be, respectively,
\begin{align}
e^\Phi\,=&\,\frac{27\left(\cos^2\xi+X^4\sin^2\xi\right)}{F^{1/2}\sin^{1/3}\xi\cos^3\xi}\,, \notag \\
C_{(0)}\,=&\,\frac{2\tilde{g}^4}{81}\rho^2-\frac{\sin^{4/3}\xi\left(3\cos^2\xi+X^4\left(2+\sin^2\xi\right)\right)}{54\left(\cos^2\xi+X^4\sin^2\xi\right)}\,.
\end{align}
The RR and NSNS two-form gauge potentials are, respectively, given by
\begin{align}
&C_{(2)}\,=\,C_{(2)}^1\, \notag \\ &=\,\frac{\tilde{g}^2R^2\rho}{3^7F}\Big[F\left(4\tilde{g}^4\rho^2-27\sin^{4/3}\xi\right)-3X^4\sec^2\xi\left(4\tilde{g}^4\rho^2(1+5\sin^2\xi)-9\cos^2\xi\sin^{4/3}\xi\right)\Big]vol_{S^2}\,, \notag \\
&B_{(2)}\,=\,C_{(2)}^2\,=\frac{2\tilde{g}^2R^2(F-X^4)\rho}{9F}vol_{S^2}\,.
\end{align}
The five-form flux vanishes identically. We checked that the uplifted solution satisfies the equations of motion of type IIB supergravity. Up to some overall factors, the uplifted solution precisely coincides with the non-Abelian T-dual of the Brandhuber-Oz solution in \cite{Lozano:2012au} which we present in appendix C.

\subsubsection{Supersymmetric $AdS_6$ black holes}

We uplift the recently obtained supersymmetric $AdS_6$ black holes of $F(4)$ gauged supergravity in \cite{Suh:2018tul} to type IIB supergravity.{\footnote{See \cite{Corbino:2018fwb} for more $AdS_2$ solutions of type IIB supergravity.}}

We first present the six-dimensional solutions of \cite{Suh:2018tul} in the conventions of \cite{Malek:2018zcz, Malek:2019ucd}. The solutions are given by the metric,
\begin{equation} \label{bhbhmet}
ds_6^2\,=\,e^{2f(r)}\left(-dt^2+dr^2\right)+e^{2g_1(r)}\left(d\theta_1^2+\sinh^2\theta_1d\phi_1^2\right)+e^{2g_2(r)}\left(d\theta_2^2+\sinh^2\theta_1d\phi_2^2\right)\,,
\end{equation}
which is asymptotically $AdS_6$ and has a horizon of $AdS_2\,\times\,\Sigma_{\frak{g}_1}\,\times\,\Sigma_{\frak{g}_2}$ with $\frak{g}_1>1$ and $\frak{g}_2>1$. Only a component of the $SU(2)$ gauge field is non-trivial,
\begin{equation} \label{bhgauge}
A^3\,=\,2\Big[a_1\cosh\theta_1{d}\phi_1+a_2\cosh\theta_2{d}\phi_2\Big]\,,
\end{equation}
where the magnetic charges, $a_1$ and $a_2$, are constant. The twist condition on the magnetic charges is
\begin{equation} \label{twist}
a_1\,=\,-\frac{k}{\lambda{g}}\,, \qquad a_2\,=\,-\frac{k}{\lambda{g}}\,,
\end{equation}
where $k\,=\,-1$ for $\frak{g}_1>1$ and $\frak{g}_2>1$ and $\lambda\,\pm\,1$. There is also a two-form field,
\begin{equation} \label{bhtwo}
B_{tr}\,=\,2\left[-\frac{2}{m^2}a_1a_2e^{\sqrt{2}\phi+2f-2g_1-2g_2}\right]\,,
\end{equation}
and the three-form field strength of the two-form field vanishes identically. There is also a non-trivial real scalar field, $X(r)\,=\,e^{\frac{\phi}{\sqrt{2}}}$, but the $U(1)$ gauge field is vanishing. To match the conventions of \cite{Malek:2018zcz, Malek:2019ucd}, there are additional overall factors of 2 to the solutions of \cite{Suh:2018tul} in \eqref{bhgauge} and \eqref{bhtwo}.

Now we perform uplifting of the supersymmetric $AdS_6$ black holes to type IIB supergravity. As it was done for the previous example, employing the uplift formula with $\mathcal{A}_\pm$ in \eqref{Afunction}, we obtain the uplifted metric,
\begin{equation} \label{IIBbh}
ds^2=\frac{1}{3\sqrt{3}}\left[\frac{F^{1/4}\cos^{3/2}\xi}{\sin^{1/6}\xi}ds_6^2+R^2\left(\frac{4\tilde{g}^4}{9}\frac{X^2\rho^2\tan^{1/2}\xi}{F^{3/4}}ds_{\tilde{S}^2}^2+\frac{F^{1/4}\tan^{1/2}\xi}{X^2}ds_\Sigma^2\right)\right]\,,
\end{equation}
where $F$ is given in \eqref{Ffunction}. The dilaton and axion fields are, respectively,
\begin{align}
e^\Phi\,=&\,\frac{27\left(\cos^2\xi+X^4\sin^2\xi\right)}{F^{1/2}\sin^{1/3}\xi\cos^3\xi}\,, \notag \\
C_{(0)}\,=&\,\frac{2\tilde{g}^4}{81}\rho^2-\frac{\sin^{4/3}\xi\left(3\cos^2\xi+X^4\left(2+\sin^2\xi\right)\right)}{54\left(\cos^2\xi+X^4\sin^2\xi\right)}\,.
\end{align}
The RR and NSNS two-form gauge potentials are, respectively, given by
\begin{align}
&C_{(2)}\,=\,C_{(2)}^1\, \notag \\ 
&=\,\frac{\tilde{g}^2c_6R^2\rho}{3^7F}\Big[F\left(4\tilde{g}^4\rho^2-27\sin^{4/3}\xi\right)-3X^4\sec^2\xi\left(2\tilde{g}^4\rho^2(1+5\sin^2\xi)-9\cos^2\xi\sin^{4/3}\xi\right)\Big]vol_{\tilde{S}^2} \notag \\
&\,\,\,\,\,\,\,\,\,\,\,\,\,\,\,\,\,\,\,\,\,\,\,\,+2\sqrt{2}c_6R\left(A^3\wedge{y}_3dk^1+A^3\wedge{k}^1Dy_3\right)+4c_6Bp^1\,, \notag \\
&B_{(2)}\,=\,C_{(2)}^2\,=\frac{2\tilde{g}^2c_6R^2(F-X^4)\rho}{9F}vol_{\tilde{S}^2}+2\sqrt{2}c_6R\left(A^3\wedge{y}_3dk^2+A^3\wedge{k}^2Dy_3\right)+4c_6Bp^2\,.
\end{align}
There is also a non-trivial self-dual five-form flux given by
\begin{equation}
F_{(5)}\,=\,F_{(2,3)}+F_{(3,2)}+F_{(4,1)}\,,
\end{equation}
where
\begin{align}
F_{(2,3)}\,=&\,\left.\frac{8\sqrt{2}c_6^2R^3}{3}\frac{\tilde{g}^2}{216}\rho\sin^{2/3}\xi\right(dA^3\wedge\epsilon_{3BC}y^BDy^C\wedge{vol}_\Sigma \notag \\
&\left.+\frac{2^83^6X^4\rho}{F\cos^4\xi}\left(y_3dA^3\wedge{p}_\alpha{dp}^\alpha-\frac{\sqrt{2}}{R}B\wedge{p}_\alpha{dk}^\alpha\right)\wedge{vol}_{\tilde{S}^2}\right)\,, \notag \\
F_{(4,1)}\,=&\,8\sqrt{2}c_6^2RX^2\left(-\gamma*_6dA^3\wedge{D}y_3+\frac{\sqrt{2}}{R}*_6B\wedge{p}_{\alpha}dp^\alpha+y_3*_6dA^3\wedge{p}_{\alpha}dk^\alpha\right)\,, \notag \\
F_{(3,2)}\,=&\,0\,.
\end{align}
The full geometry is interpolating between the asymptotic $AdS_6$ fixed point, which is the non-Abelian T-dual of the Brandhuber-Oz solution, and the near horizon geometry, $AdS_2\times\Sigma_{\frak{g}_1}\times\Sigma_{\frak{g}_2}\times\tilde{S}^2\times\Sigma$. By employing the supersymmetry equations of $F(4)$ gauged supergravity, we explicitly checked that the uplifted solution solves the equations of motion. For the Einstein equations, due to the complexity of the solution, we checked them for numerous specific numerical values of the coordinates on the Riemann surface, $(\rho,\,\xi)$. We present the supersymmetry equations of $F(4)$ gauged supergravity in appendix A and the equations of motion of type IIB supergravity in appendix B.2.

\subsection{Holographic entanglement entropy}

The black hole solution interpolates between the boundary, $AdS_6\times\tilde{S}^2\times\Sigma$, and the horizon, $AdS_2\times\Sigma_{\frak{g}_1}\times\Sigma_{\frak{g}_2}\times\tilde{S}^2\times\Sigma$. The holographic entanglement entropy, \cite{Ryu:2006bv, Ryu:2006ef}, on the boundary was calculated in \cite{Lozano:2013oma}. In this section, we calculate the holographic entanglement entropy on the horizon.

For the gravitational theory with a dilaton field, the holographic entanglement entropy is given by, \cite{Klebanov:2007ws}, 
\begin{equation}
S_{\text{EE}}\,=\,\frac{1}{4G_N^{(10)}}\int{d^8x}e^{-2\Phi}\sqrt{\gamma}\,,
\end{equation}
where $\gamma$ is determinant of the induced metric of co-dimension two entangling surface in string frame. We compute the holographic entanglement entropy on the horizon of $AdS_2\times\Sigma_{\frak{g}_1}\times\Sigma_{\frak{g}_2}\times\tilde{S}^2\times\Sigma$. Unlike computing entanglement entropy on the asymptotically $AdS_6$ boundary in \cite{Jafferis:2012iv}, the minimal surface is just a point for $AdS_2$, \cite{Azeyanagi:2007bj}. In this case, the holographic entanglement entropy computes the entanglement between CFTs on two boundaries of $AdS_2$ space. Then, the holographic entanglement entropy is obtained to be
\begin{align}
S_{\text{EE}}\,=&\,\frac{1}{4G_N^{(10)}}\int{d}^8x\frac{2^4}{3^{10}}c_6^4R^4\tilde{g}^6e^{2g_1+2g_2}\rho^2\cos^2\xi\sin^{1/3}\xi\sinh\theta_1\sinh\theta_2\sin\theta \notag \\
=&\,\frac{1}{4G_N^{(10)}}\frac{2^4}{3^{10}}c_6^4R^4\tilde{g}^6e^{2g_1}vol_{\Sigma_{\mathfrak{g}_1}}e^{2g_2}vol_{\Sigma_{\mathfrak{g}_2}}vol_{\tilde{S}^2}\int_0^Rd\rho\rho^2\int_0^{\pi/2}d\xi\cos^2\xi\sin^{1/3}\xi \notag \\
=&\,\frac{2^4}{3^{12}5\pi^3}c_6^4R^7\tilde{g}^6(\mathfrak{g}_1-1)(\mathfrak{g}_2-1)\,,
\end{align}
where we use $e^{2g_1+2g_2}\,=\,2/3^3$ from \eqref{ads2sol}. If the field theory dual of the non-Abelian T-dual of the Brandhuber-Oz solution could be identified, it would be interesting to compare the holographic entanglement entropy we obtain here with the microscopic results.

\medskip
\bigskip
\leftline{\bf Acknowledgements}
\noindent We would like to thank Emanuel Malek for explaining his work, and for very important comments on a preprint  which corrected an error and improved the preprint. We also thank Tatsuma Nishioka for helpful communications. This research was supported by the National Research Foundation of Korea under the grant NRF-2019R1I1A1A01060811.

\appendix
\section{The supersymmetry equations of $F(4)$ gauged supergravity}
\renewcommand{\theequation}{A.\arabic{equation}}
\setcounter{equation}{0} 

We present the supersymmetry equations of $AdS_6$ black holes from $F(4)$ gauged supergravity, \cite{Suh:2018tul}, in the conventions of \cite{Romans:1985tw, Suh:2018tul},
\begin{align} \label{susy111}
f'e^{-f}\,=&\,-\frac{1}{4\sqrt{2}}\left(ge^{\frac{\phi}{\sqrt{2}}}+me^{-\frac{3\phi}{\sqrt{2}}}\right)-\frac{\lambda}{2\sqrt{2}}e^{-\frac{\phi}{\sqrt{2}}}\left(a_1e^{-2g_1}+a_2e^{-2g_2}\right)-\frac{3}{\sqrt{2}m}a_1a_2e^{\frac{\phi}{\sqrt{2}}-2g_1-2g_2}\,, \notag \\ 
g_1'e^{-f}\,=&\,-\frac{1}{4\sqrt{2}}\left(ge^{\frac{\phi}{\sqrt{2}}}+me^{-\frac{3\phi}{\sqrt{2}}}\right)+\frac{\lambda}{2\sqrt{2}}e^{-\frac{\phi}{\sqrt{2}}}\left(3a_1e^{-2g_1}-a_2e^{-2g_2}\right)+\frac{1}{\sqrt{2}m}a_1a_2e^{\frac{\phi}{\sqrt{2}}-2g_1-2g_2}\,, \notag \\ 
g_2'e^{-f}\,=&\,-\frac{1}{4\sqrt{2}}\left(ge^{\frac{\phi}{\sqrt{2}}}+me^{-\frac{3\phi}{\sqrt{2}}}\right)+\frac{\lambda}{2\sqrt{2}}e^{-\frac{\phi}{\sqrt{2}}}\left(3a_2e^{-2g_2}-a_1e^{-2g_1}\right)+\frac{1}{\sqrt{2}m}a_1a_2e^{\frac{\phi}{\sqrt{2}}-2g_1-2g_2}\,, \notag \\ 
\frac{1}{\sqrt{2}}\phi'e^{-f}\,=&\,\frac{1}{4\sqrt{2}}\left(ge^{\frac{\phi}{\sqrt{2}}}-3me^{-\frac{3\phi}{\sqrt{2}}}\right)+\frac{\lambda}{2\sqrt{2}}e^{-\frac{\phi}{\sqrt{2}}}\left(a_1e^{-2g_1}+a_2e^{-2g_2}\right)-\frac{1}{\sqrt{2}m}a_1a_2e^{\frac{\phi}{\sqrt{2}}-2g_1-2g_2}\,.
\end{align}

\section{The equations of motion of type II supergravity}
\renewcommand{\theequation}{B.\arabic{equation}}
\setcounter{equation}{0} 

We present the equations of motion of type II supergravity in Einstein frame. The metric in string frame is obtained by
\begin{equation}
g_{MN}^{\text{string}}\,=\,e^{\Phi/2}g_{MN}^{\text{Einstein}}\,.
\end{equation}
We also define
\begin{equation}
|F_{(p)}|^2\,=\,\frac{1}{p!}F_{(p)M_1{\cdots}M_p}F_{(p)}\,^{M_1{\cdots}M_p}\,, \qquad |F_{(p)}|_{MN}^2\,=\,\frac{1}{(p-1)!}F_{(p)MM_1{\cdots}M_{p-1}}F_{(p)N}\,^{M_1{\cdots}M_{p-1}}\,.
\end{equation}

\subsection{Massive type IIA supergravity}
\renewcommand{\theequation}{B.\arabic{equation}}
\setcounter{equation}{0} 

The field content of massive type IIA supergravity is the metric, $g_{MN}$, the dilaton field, $\Phi$, the NSNS two-form gauge potential, $B_{(2)}$, and the RR one- and three-form gauge potentials, $C_{(1)}$ and $C_{(3)}$, respectively, \cite{Romans:1985tz}. We follow the conventions of \cite{Cvetic:1999un}. The fluxes are defined by
\begin{equation}
F_{(2)}\,=\,dC_{(1)}+\frak{m}B_{(2)}\,, \qquad H_{(3)}\,=\,dB_{(2)}\,, \qquad F_{(4)}\,=\,dC_{(3)}+C_{(1)}\wedge{d}B_{(2)}+\frac{\frak{m}}{2}B_{(2)}\wedge{B}_{(2)}\,,
\end{equation}
where $\frak{m}$ is the mass parameter of $B_{(2)}$, known as the Romans' mass, and is related to the $SU(2)$ coupling of $F(4)$ gauged supergravity by
\begin{equation}
\frak{m}\,=\,\frac{\sqrt{2}}{3}g\,.
\end{equation}
The equations of motion are given by
\begin{align}
R_{MN}&-\frac{1}{2}\partial_M\Phi\partial_N\Phi-\frac{1}{16}\frak{m}^2e^{\frac{5}{2}\Phi}g_{MN}-\frac{1}{2}e^{\frac{3}{2}\Phi}\left(|F_{(2)}|_{MN}^2-\frac{1}{8}|F_{(2)}|^2g_{MN}\right) \notag \\
&-\frac{1}{2}e^{-\Phi}\left(|H_{(3)}|_{MN}^2-\frac{1}{4}|H_{(3)}|^2g_{MN}\right)-\frac{1}{2}e^{\frac{1}{2}\Phi}\left(|F_{(4)}|^2_{MN}-\frac{3}{8}|F_{(4)}|^2g_{MN}\right)\,=\,0\,,
\end{align}
\begin{equation}
\frac{1}{\sqrt{-g}}\partial_M\left(\sqrt{-g}g^{MN}\partial_N\Phi\right)-\frac{5}{4}\frak{m}^2e^{\frac{5}{2}\Phi}-\frac{3}{4}e^{\frac{3}{2}\Phi}|F_{(2)}|^2+\frac{1}{2}e^{-\Phi}|H_{(3)}|^2-\frac{1}{4}e^{\frac{1}{2}\Phi}|F_{(4)}|^2\,=\,0\,,
\end{equation}
\begin{align}
d\left(e^{-\Phi}*H_{(3)}\right)+\frac{1}{2}F_{(4)}\wedge{F}_{(4)}+\mathfrak{m}e^{3\Phi/2}*F_{(2)}+e^{\Phi/2}*F_{(4)}\wedge{F}_{(2)}\,=\,0\, \notag \\
d\left(e^{3\Phi/2}*F_{(2)}\right)+e^{\Phi/2}*F_{(4)}\wedge{H}_{(3)}\,, \notag \\
d\left(e^{\Phi/2}*F_{(4)}\right)+H_{(3)}\wedge{F}_{(4)}\,=\,0\,,
\end{align}
and the Bianchi identities are
\begin{equation}
dF_{(2)}\,=\,\frak{m}H_{(3)}\,, \qquad dH_{(3)}\,=\,0\,, \qquad dF_{(4)}\,=\,F_{(2)}\wedge{H}_{(3)}\,.
\end{equation}

\subsection{Type IIB supergravity}
\renewcommand{\theequation}{B.\arabic{equation}}
\setcounter{equation}{0} 

The field content of type IIB supergravity is the metric, $g_{MN}$, the dilaton and the axion fields, $\Phi$ and $C_{(0)}$, the NSNS and RR two-form gauge potential, $B_{(2)}$ and $C_{(2)}$, and the RR four-form gauge potential, $C_{(4)}$, respectively, \cite{Schwarz:1983qr, Howe:1983sra}. We follow the conventions of \cite{Bobev:2018hbq}. The fluxes are
\begin{equation}
F_{(1)}\,=\,dC_{(0)}\,, \qquad H_{(3)}\,=\,dB_{(2)}\,, \qquad F_{(3)}\,=\,dC_{(2)}-C_{(0)}H_{(3)}\, \qquad F_{(5)}\,=\,\mathcal{F}_{(5)}+*_{10}\mathcal{F}_{(5)}\,,
\end{equation}
where
\begin{equation}
\mathcal{F}_{(5)}\,=\,dC_{(4)}-\frac{1}{2}\left(C_{(2)}\wedge{H}_{(3)}-B_{(2)}\wedge{dC}_{(2)}\right)\,.
\end{equation}
The equations of motion are given by
\begin{align}
R_{MN}&-\frac{1}{2}\partial_M\Phi\partial_N\Phi-\frac{1}{2}e^{2\Phi}|F_{(1)}|_{MN}^2-\frac{1}{2}e^\Phi\left(|F_{(3)}|_{MN}^2-\frac{1}{4}|F_{(3)}|^2g_{MN}\right) \notag \\
&-\frac{1}{2}e^{-\Phi}\left(|H_{(3)}|_{MN}^2-\frac{1}{4}|H_{(3)}|^2g_{MN}\right)-\frac{1}{4}e^{-\Phi}\left(|F_{(5)}|^2_{MN}-\frac{1}{2}|F_{(5)}|^2g_{MN}\right)\,=\,0\,,
\end{align}
\begin{equation}
\frac{1}{\sqrt{-g}}\partial_M\left(\sqrt{-g}g^{MN}\partial_N\Phi\right)-e^{2\Phi}|F_{(1)}|^2+\frac{1}{2}e^{-\Phi}|H_{(3)}|^2-\frac{1}{2}e^\Phi|F_{(3)}|^2\,=\,0\,,
\end{equation}
\begin{align}
d\left(e^{-\Phi}*H_{(3)}\right)-F_{(3)}\wedge{F}_{(5)}-e^\Phi{F}_{(1)}\wedge*F_{(3)}\,=\,0\,, \notag \\
d\left(e^{2\Phi}*F_{(1)}\right)+e^\Phi{H}_{(3)}\wedge*F_{(3)}\,=\,0\,, \notag \\
d\left(e^\Phi*F_{(3)}\right)+H_{(3)}\wedge{F}_{(5)}\,=\,0\,,
\end{align}
and the Bianchi identities are
\begin{equation}
dF_{(1)}\,=\,0\,, \qquad dH_{(3)}\,=\,0\,, \qquad dF_{(3)}-H_{(3)}\wedge{F}_{(1)}\,=\,0\, \qquad dF_{(5)}-H_{(3)}\wedge{F}_{(3)}\,=\,0\,.
\end{equation}

\section{The non-Abelian T-dual of the Brandhuber-Oz solution}
\renewcommand{\theequation}{C.\arabic{equation}}
\setcounter{equation}{0} 

We present the non-Abelian T-dual of the Brandhuber-Oz solution, \cite{Lozano:2012au}, as it appeared in \cite{Jeong:2013jfc}.{\footnote {The scalar field in \cite{Jeong:2013jfc} is related to the scalar field in this work by $X^{\text{there}}\,=\,1/{X^{\text{here}}}$. We use our $X$ in this appendix. We also rescaled the metric in string frame in \cite{Jeong:2013jfc} to the metric in Einstein frame.} The solution is consist of the metric,
\begin{equation}
ds^2\,=\,e^{-\Phi/2}\left[\tilde{\Delta}^{1/2}\sin^{-1/3}{\xi}ds_{AdS_6}^2+\frac{\rho^2e^{2A}}{\rho^2+e^{4A}}ds_{S^2}^2+e^{-2A}d\rho^2+\frac{2}{\tilde{g}^2}X^{-2}\tilde{\Delta}^{1/2}\sin^{-1/3}{\xi}d\xi^2\right]\,,
\end{equation}
the dilaton and axion fields, respectively,
\begin{align}
e^{\Phi}\,=&\,\frac{X\tilde{\Delta}^{1/4}}{e^A\sqrt{\rho^2+e^{4A}}\sin^{5/6}\xi}\,, \notag \\
F_{(1)}\,=&\,-G_1+\tilde{m}{\rho}d\rho\,,
\end{align}
and the NSNS two-form gauge potential and the RR three-form flux are, respectively,
\begin{align}
B_{(2)}\,=&\,-\frac{\rho^3}{\rho^2+e^{4A}}vol_{S^2}\,, \notag \\
F_{(3)}\,=&\,\frac{\rho^2}{\rho^2+e^{4A}}\left(\rho{G}_1+\tilde{m}e^{4A}d\rho\right)\wedge{vol}_{S^2}\,,
\end{align}
where we define
\begin{align}
\tilde{\Delta}\,=&\,\cos^2\xi+X^4\sin^2\xi\,, \notag \\
e^A\,=&\,\frac{X\cos\xi}{\sqrt{2}\tilde{g}\tilde{\Delta}^{1/4}\sin^{1/6}\xi}\,, \notag \\
G_1\,=&\,-\frac{1}{12}UX^2\tilde{\Delta}^{-2}\sin^{1/3}\xi\cos^3{\xi}d\xi-\frac{1}{4}X^3\tilde{\Delta}^{-2}\sin^{4/3}\xi\cos^4{\xi}dX\,, \notag \\
U\,=&\,X^6\sin^2\xi-3X^{-2}\cos^2\xi+4X^2\cos^2\xi-6X^2\,.
\end{align}
The five-form flux vanishes identically.

In order to match the non-Abelian T-dual of the Brandhuber-Oz solution with the uplifted $AdS_6$ fixed point in section 3.2.1, the uplifted solution should be rescaled by
\begin{align}
\tilde{g}_{\mu\nu}\,=\,\frac{1}{\tilde{m}^{3/2}}g_{\mu\nu}\,, \qquad e^{\tilde{\Phi}}\,=\,\tilde{m}^3e^\Phi\,, \qquad \tilde{C}_{(0)}\,=\,\frac{1}{\tilde{m}^3}C_{(0)}\,, \notag \\
\tilde{C}_{(2)}\,=\,\frac{1}{\tilde{m}^3}C_{(2)}\,, \qquad \tilde{B}_{(2)}\,=\,B_{(2)}\,, \qquad \tilde{F}_{(5)}\,=\,\frac{1}{\tilde{m}^3}F_{(5)}
\end{align}
and take, $e.g.$, in \cite{Jeong:2013jfc},
\begin{equation}
\tilde{m}\,=\,\frac{\sqrt{2}}{3}\tilde{g}\,=\,\frac{2}{3}\,,
\end{equation}
where the tilded ones are the rescaled fields.



\end{document}